\documentclass[fleqn,12pt]{wlscirep}
\usepackage[utf8]{inputenc}
\usepackage[T1]{fontenc}
\usepackage{caption}
\usepackage{subcaption}
\usepackage{graphicx}
\usepackage{multibib}
\usepackage{soul}
\usepackage{color}
\usepackage{ulem}
\usepackage{epstopdf}
\AppendGraphicsExtensions{.tiff}

\addto\captionsenglish{}

\newcommand{\msun}{M$_\odot$}
\newcommand{\al}{$\alpha$}
\newcommand{\kpckms}{kpc\ km\ s$^{-1}$}
\newcommand{\jz}{$J_z$}

\newcommand{\nat}{Nature}
\newcommand{\apj}{Astrophys. J.}
\newcommand{\aap}{Astron. Astrophys.}
\newcommand{\aj}{Astron . J.}
\newcommand{\apjl}{Astrophys. J. Let.}
\newcommand{\apjs}{Astrophys. Js}
\newcommand{\araa}{Annu. Rev. Astron. Astrophys.}
\newcommand{\mnras}{Mon. Not. R. Astron. Soc.}
\newcommand{\pasp}{Publ. Astron. Soc. Pac.}

\title{Stellar Initial Mass Function Varies with Metallicities and Time}

\author[1,2,3]{Jiadong Li}
\author[1,2,3,*]{Chao Liu}
\author[4,5]{Zhi-Yu Zhang}
\author[1]{Hao Tian}
\author[6,7]{Xiaoting Fu}
\author[1]{Jiao Li}
\author[4,5]{Zhi-Qiang Yan}

\affil[1]{Key Lab of Space Astronomy and Technology, National Astronomical Observatories, Beijing, 100101, China}
\affil[2]{Institute for Frontiers in Astronomy and Astrophysics, Beijing Normal University, Beijing, 102206, China}
\affil[3]{University of Chinese Academy of Sciences, Beijing, 100049, China}
\affil[4]{School of Astronomy and Space Science, Nanjing University, Nanjing, 210000, China}
\affil[5]{Key Laboratory of Modern Astronomy and Astrophysics, Nanjing University, Ministry of Education, Nanjing 210093, China}
\affil[6]{Purple Mountain Observatory, Chinese Academy of Sciences, Nanjing, 210023, China}
\affil[7]{The Kavli Institute for Astronomy and Astrophysics at Peking University, Beijing, 100871, China}

\affil[*]{corresponding author, liuchao@nao.cas.cn}
\begin{abstract}
\textbf{
Most structural and evolutionary properties of galaxies strongly rely on the stellar initial mass function (IMF), namely the distribution of the stellar mass formed in each episode of star formation$^{1-4}$. As the IMF shapes the stellar population in all stellar systems, it turns out to become one of the most fundamental concepts of modern astronomy. Both constant and variable IMFs across different environments have been claimed despite a large number of theoretical $^{5-7}$ and observational efforts $^{8,10,12,14-16}$. However, the measurement of the IMF in Galactic stellar populations has been limited by the relatively small number of photometrically observed stars, leading to high uncertainties $^{12-16}$. Here we report a star-counting result based on $\sim$93,000 spectroscopically observed M-dwarf stars, an order of magnitude more than previous studies, in the 100--300 parsec (pc) Solar neighbourhood. We find unambiguous evidence of a variable IMF that depends on both metallicity and stellar age. Specifically, the stellar population formed at the early time contains fewer low-mass stars compared to the canonical IMF, independent of stellar metallicities. In present days, on the other hand, the proportion of low-mass stars increases with stellar metallicity. The variable abundance of low-mass stars in our Milky Way establishes a powerful benchmark for models of star formation and can heavily impact results in Galactic chemical enrichment modelling, mass estimation of galaxies, and planet formation efficiency.
}
\end{abstract}
\begin{document}

% \flushbottom
\maketitle

\thispagestyle{empty}

% \noindent Please note: Abbreviations should be introduced at the first mention in the main text – no abbreviations lists. The suggested structure of main text (not enforced) is provided below.

% \section*{Introduction}

\begin{figure*}[ht!]
\includegraphics[width=\linewidth]{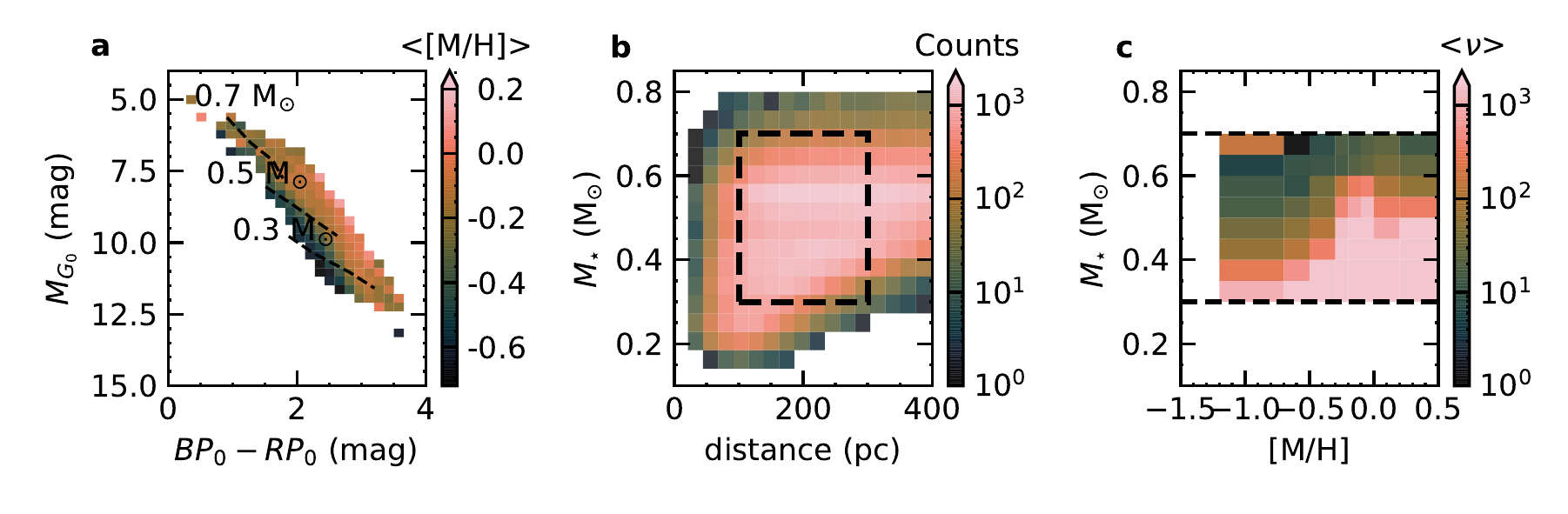}
  \hspace*{\fill} 
\caption{\textbf{| The M dwarf star sample in the solar neighbourhood. a,} Colour-magnitude distribution of the samples of M-type stars. The colour bar indicates the median stellar metallicity, [M/H], in each colour--magnitude bin. The dashed lines indicate the stellar mass from PARSEC stellar models. \textbf{b,} The stellar counts as a function of stellar mass and distance to the Sun. We chose stars inside the black dashed box, within which the coverage of the stellar mass is roughly uniform along the distance. Hence, the selected stars are barely affected by any volume selection effect. \textbf{c,} The stellar number density after selection effect correction as a function of stellar mass and metallicity, [M/H]. The two horizontal dashed lines indicate the selected stellar mass range, $0.3$ \msun $<$M$^\star$ $<$ $0.7$\,\msun.}\label{fig:sample} 
\end{figure*}

\begin{figure*}[hbt!]
\centering
\includegraphics[width=0.99\linewidth]{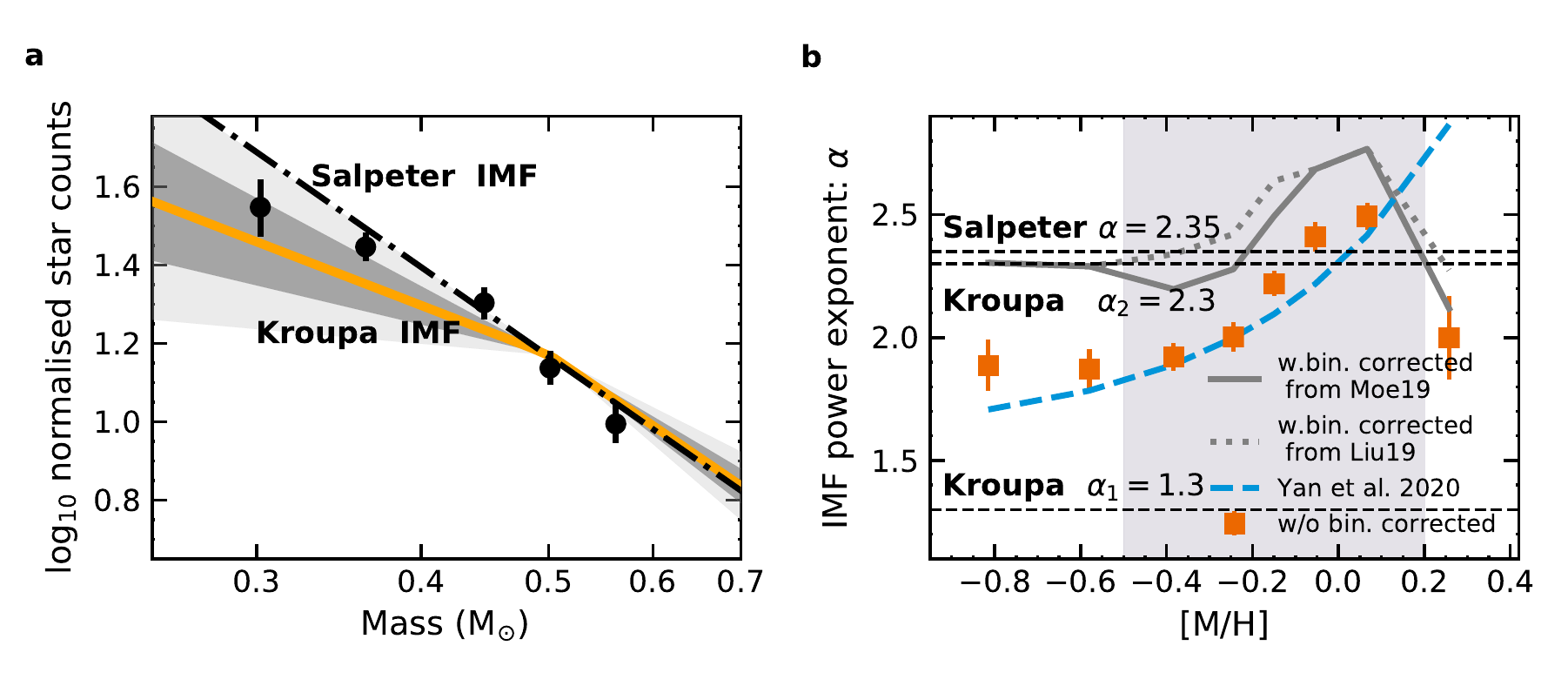} \caption{\textbf{| Stellar inital mass function variation revealed by our M dwarf star sample. a,} IMFs predicted by the posterior of the Bayesian models. All IMFs are normalised at 0.5\,\msun\ for comparison. The yellow solid line indicates the Kroupa IMF with the power exponents $\alpha_1=1.3$ at $<0.5$\msun\ and $\alpha_2=2.3$ at $>0.5$\msun\,. The dimgrey shaded region represents the 1-$\sigma$ uncertainty of Kroupa IMF, and the lightgrey region denotes the 2-$\sigma$ uncertainty region. The dash-dotted line indicates the Salpeter IMF with the power exponent of 2.35. \textbf{b,} The IMF power exponent, \al, as a function of stellar metallicity. The filled rectangles display \al\ at different [M/H] bins. The grey solid and dotted lines represent \al--[M/H] relation, after correcting the binary star fractions\cite{liu2019,2019ApJ...875...61M}. The blue dashed line denotes the IMF variation used in literature\cite{2020A&A...637A..68Y}. The vertical shaded region emphasizes the metallicity range of $-0.5 < {\rm [M/H]} < +0.2$, in which \al\ tightly increases with [M/H]. The error bars denote the standard deviations of IMF slopes. Error bars represent 1$\sigma$ uncertainty.}
\label{fig:alpha}
\end{figure*}

Direct star counting, which provides the most direct and least model-dependent evidence\cite{1993MNRAS.262..545K}, is essential to verify the IMF variation in our Milky Way. The classical studies apply mass-luminosity relation to derive IMF from observed luminosity function\cite{1997MNRAS.287..402K} and make the canonical IMF as a benchmark\cite{2002Sci...295...82K}. Recent large surveys such as Gaia and Large Sky Area Multi-Object Fiber Spectroscopic Telescope (LAMOST) offers a new opportunity to reassess the IMF with the star-counting technique.  
  
We focus on the low-mass stars in the Milky Way by counting field stars with a stellar mass range of $0.3\sim0.7$\,\msun, which are mostly M-dwarf stars. Their lifespans are longer than the Universe, so none has evolved out of the main sequence. Therefore, their current mass distribution, i.e., the present-day mass function, is essentially identical to their IMF, irrespective of the star formation history.
 
First, we selected $\sim$ 300,000 M-dwarf stars with effective temperature ($T_{\rm eff}$) and metallicity ([M/H]) obtained from a catalogue based on LAMOST and Gaia data\cite{jdli21}. We derived their stellar masses with the typical uncertainty of 0.01 \msun\ by comparing the stellar parameters with the PAdova and TRieste Stellar Evolution Code (PARSEC) stellar model (see the Methods for details). Fig.1 (a) displays the colour-magnitude diagram of the stellar samples, with theoretical grids of stellar mass overlaid. 
Second, we chose stars with distance to the Sun from 100 to 300 pc and stellar mass between $0.3$ and $0.7$\,\msun\ to avoid the selection effect of the stellar mass at different distances (see Fig.1 b). The mass coverage is also essentially constant in different metallicities, as displayed in Fig.~1 (c). 
Third, we measured the stellar density of sub-samples in each metallicity-mass bin after correcting the selection effect\cite{liu2017} (see Methods) of the LAMOST spectroscopic survey by comparing it with Two Micron All Sky Survey (2MASS) photometric data.
Last, we corrected the bias from binary populations based on the observed binary information from the literature\cite{liu2019,2019ApJ...875...61M} (see Methods).

%With the final samples after all corrections, it is still necessary to consider the stellar density profile of the Milky Way disk.

To count stars with different masses in the Solar vicinity, we assume that the IMF of a given metallicity should be the same everywhere in the studied volume. We then adopted a hierarchical Bayesian model for deriving the posterior probability density distribution of \al, which is the power-law exponent of the IMF for stars with 0.3--0.7 $M_{\odot}$, from the stellar density profiles with different stellar masses. In this model, we adopted a single exponential vertical stellar density profile, up to the largest distance of our selected samples, 300 pc, from the Sun.
 
We find our results of all metallicity stars are consistent with canonical IMFs within 1-$\sigma$ uncertainty as shown in Fig~2a. And our result shows that the stellar IMF is obviously not universal (see Fig~2b). The slope of IMFs changes around the canonical IMF and Salpeter IMF with metallicities. As the [M/H] increases from $-0.8$ to $+0.1$, the IMF slope continuously moves from 1.9 to 2.5. This trend is quantitatively displayed in Fig~2 (a), in which \al\ is directly associated with metallicity. At the same time, \al\ does not vary with metallicity when [M/H]$<-0.3$. It is tightly correlated with metallicity at $-0.5<$[M/H]$\le+0.1$. The most bottom-heavy (biased to low-mass stars) IMF occurs at [M/H]$\sim+0.1$.
 
When [M/H] changes from $-0.8$ to $+0.1$, \al\ increases by $\sim$0.6. Such an amplitude of variation is in well agreement with empirical predictions\cite{2002Sci...295...82K} and models\cite{2020A&A...637A..68Y}, which suggest that the $\mathrm{d}{\alpha} / \mathrm{d}{\rm [M/H]} \simeq 0.5$ for $m < 0.7 M_{\odot}$. The \al\,--[M/H] trend is similar to the variation of the IMF adopted in literatures\cite{2020A&A...637A..68Y}, which was derived from resolved star counts as well as globular clusters, ultra-compact dwarf galaxies, ultra-faint dwarf galaxies and massive elliptical galaxies. It also reconciles other observational results, e.g., metal-poor thick-disc stars\cite{2001A&A...373..886R} and ultra-faint dwarf galaxies \cite{2013ApJ...771...29G} have bottom-light IMFs. 

% It also reconciles other observational results, e.g., younger clusters with higher [M/H] have more bottom-heavy IMFs\cite{2013pss5.book..115K} and metal-poor old thick-disc stars ([Fe/H]$=-0.6$) and Galactic globular clusters ([Fe/H]$<-1$) have more bottom-light IMFs\cite{2001A&A...373..886R,2002Sci...295...82K}.
 
However, the sharp turnoff between [M/H]$=+0.1$ and [M/H]$=+0.25$\,dex (\al\ drops from $2.50 \pm 0.06$ to $2.00 \pm 0.17$) deviates from the increasing trend of \al--[M/H] in the regime of -0.5$<$[M/H]$\le+0.1$, which is significant at $\sim$ 2.8$\sigma$ as shown in Extended Data Fig.1. The majority of stars with [M/H]$>+0.1$ are likely migrated from a few kpc closer to the Galactic centre than the Sun in the inner Galactic disc\cite{2015MNRAS.447.3526K}. Stars with [M/H]$\sim+0.25$ are likely from relatively ancient populations, as found in previous studies\cite{2015MNRAS.447.3526K}. A reasonable inference is that the IMF is not only related to metallicity but also to the star-forming environment\cite{2018A&A...620A..39J} where the migrated stars were born. For instance, they may be formed under a high star formation rate regions in the inner Galactic disc, which may produce top-heavy IMFs\cite{2010ApJ...708..834B}. 

\begin{figure*}[ht!]
\centering
\includegraphics[width=0.9\linewidth]{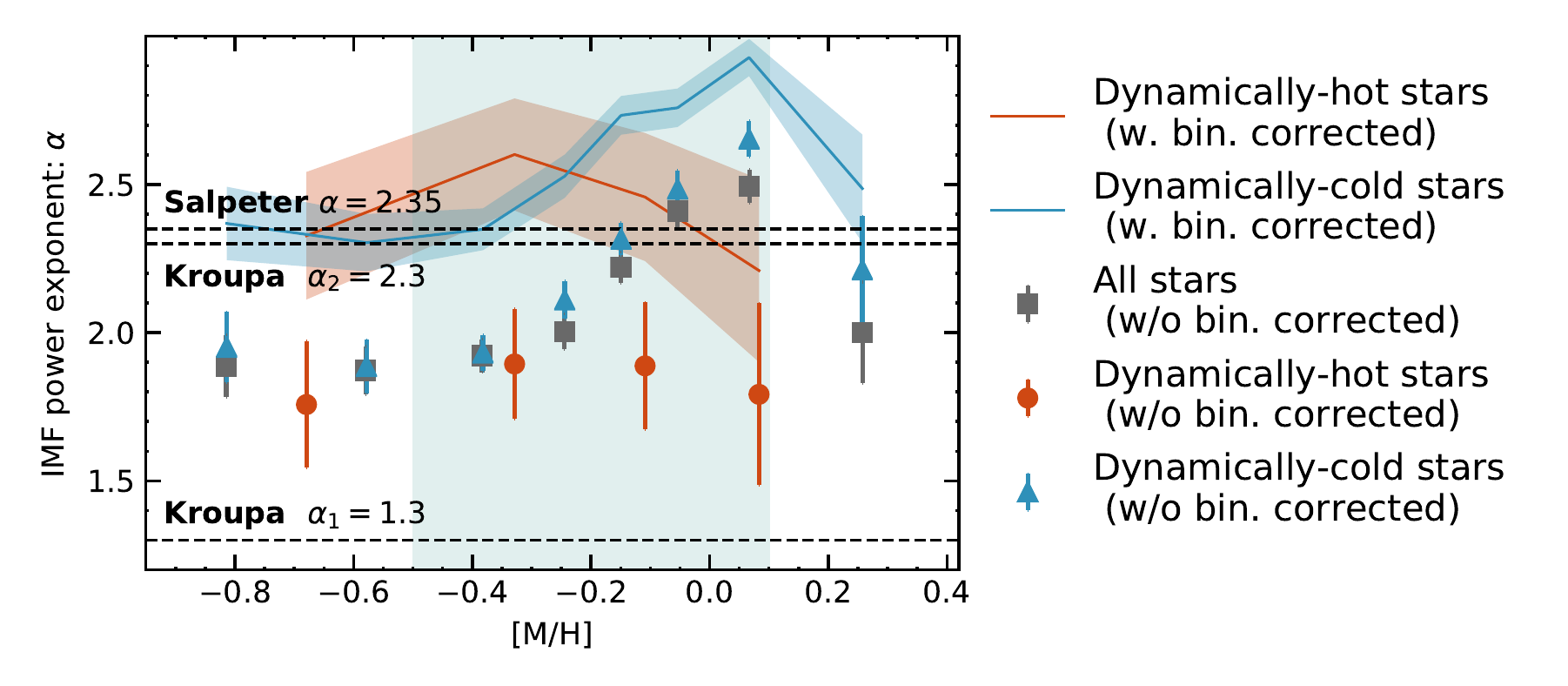}
\caption{\textbf{| The IMF power exponent as a function of stellar metallicity for different groups.} Dynamically-hot stars (\jz$>20$\,\kpckms) are denoted as red filled circles, and dynamically-cold stars (\jz$<10$\,\kpckms) are shown as blue triangles, respectively. The red and blue solid lines represent \al, after correcting their binary star fractions, of dynamically-hot and dynamically-cold stars, respectively\cite{liu2019}. The binary fractions of dynamically-hot stars are set as a factor of $1.5$ larger than those of the dynamically-cold stars at same metallicity. The red and blue shades represent their 1-$\sigma$ uncertainty regions. Error bars represent 1$\sigma$ uncertainty.}
\label{fig:3}
\end{figure*}

We further examine the IMF variation against their dynamic property of vertical action (\jz), which could roughly trace stellar ages. Stars with larger \jz\ have experienced more scattering with molecular clouds and thus are more likely old, while those with small \jz\ are not necessarily young since old stars may also have small \jz\ if they were not effectively scattered over a long time\cite{2019ApJ...878...21T}. Therefore, we split the samples into dynamically-hot and -cold groups with \jz$>20$\,\kpckms\ and \jz$<10$\,\kpckms (see Methods), respectively (see Fig~3). 

When [M/H] $\le-0.5$\,dex, where most stars are older than four billion years (see Methods), both dynamically-hot and -cold groups show constant power-law indices of the IMFs (Fig. 3). When [M/H]$\geq-0.4$\,dex, however, their variation trends of \al\ split out, between dynamically-hot and -cold groups. The variation slope of dynamically-cold star is $\mathrm{d}{\alpha} / \mathrm{d}{\rm [M/H]} = 1.17 \pm 0.10$ with $-0.5<{\rm [M/H]}\le+0.1$. In the same [M/H] regime, the slope is $\mathrm{d}{\alpha} / \mathrm{d}{\rm [M/H]} = -0.17 \pm 0.18$ for dynamically-hot stars.
% \lc{We need to fit the slopes of \al\ separately for cold and hot groups and show that the slopes are significantly different.}

While dynamically-cold stars follow the IMF variation of the whole sample, dynamically-hot stars show a flat \al\ trend with metallicity (see Fig~3). Although the sample contains a limited number of stars with \jz$>20$\,\kpckms\ at [M/H]$=+0.25$, their average \jz\ biases to larger values, which is consistent with the argument that they are relatively-old migrated populations (see Methods). Therefore, it is not surprising that \al\ drops to a value similar to the dynamically-hot and metal-poor stars. 

The fraction of binary stars may also bias the IMF determination. Therefore, we corrected the \al\ estimates by considering scenarios of a general binary fraction from observations\cite{liu2019,2019ApJ...875...61M}, a high binary fraction for dynamically-hot stars (45\%), and an extreme case with all stars in binary systems (see Methods). They all end up with the similar \al$-$[M/H] trend with increased \al\ by 0.2--0.4 after the binary correction. 

Our findings present a robust observational confirmation of the variability of IMF in the Milky Way, based on star counting in the Solar neighbourhood. The dependencies on stellar metallicity and age indicate that exotic physical environments, such as variable Jeans mass \cite{1998MNRAS.301..569L}, turbulence \cite{2002ApJ...576..870P}, cosmic rays \cite{2011MNRAS.414.1705P}, and the cosmic microwave background \cite{2016RSOS....360025Z}, etc., may systematically influence star-forming processes and shape the IMF. The IMF variation also calls for an extensive revision of star formation rates and stellar mass in the Milky Way and external galaxies, especially those with extreme physical conditions and in the early Universe. Future studies, such as the large field surveys from the forthcoming Chinese Space Station Telescope (CSST), will provide insights into the IMF variation in larger volumes, lower stellar mass and probably more stellar clusters.

\section*{Methods}
\subsection*{Sample selection} 

We select the sample from a public catalogue of $\sim$ 300,000 M-dwarf stars\cite{jdli21}, which have precise spectroscopic stellar parameters estimated from LAMOST low-resolution spectra (R$\sim 1800$)\cite{2012RAA....12..723Z,2012RAA....12..735D}. The stellar parameters of the LAMOST M dwarfs are trained based on the Sloan Digital Sky Survey Apache Point Observatory Galactic Evolution Experiment (SDSS/APOGEE)\cite{2017AJ....154...94M} Data Release 16\cite{2020AJ....160..120J} using SLAM algorithm\cite{2020ApJS..246....9Z}.
Typical uncertainties of effective temperature (${T_{\rm eff}}$) and stellar metallicity ([M/H]) are 60\,K and 0.12 dex, respectively. The accuracy of metallicity provided in this catalogue has been well validated in both open clusters and wide binaries\cite{jdli21}. We also collected the line-of-sight velocities measured by the LAMOST pipeline\cite{2014AJ....147...33Y} and the proper motions from Gaia DR2\cite{gaia2018b} of these M-dwarf stars. We further crossmatch with the catalogue of 2MASS\cite{2006AJ....131.1163S} by angular distances within 5$''$. The 3-D dust reddening maps from \textit{Bayestar}\cite{2019ApJ...887...93G} are used to estimate the visual extinction $A_{\rm V}$ for each star, and $A_{\rm K_s}$ is further estimated from $A_{\rm V}$ using the extinction factor from literature \cite{2019ApJ...877..116W}. Then, we adopt the Bayesian distance ($r_{\rm est}$) derived from the parallax of Gaia\cite{2018AJ....156...58B} and measured the absolute magnitude ($M_{K_s}$) at the $K_s$ band of 2MASS.

\subsection*{Stellar masses}\label{subsec:mass}
The stellar mass of each star is derived by comparing their ${\rm T_{\rm eff}}$ and $M_{K_s}$ with PARSEC isochrones\cite{2012MNRAS.427..127B,2014MNRAS.444.2525C} using XGboost algorithms\cite{chen2016xgboost}. The specific training procedure uses the root mean square error as the loss function. A 10-fold cross-validation has been conducted, and the overall uncertainty introduced by the algorithm is $\sim0.03$\,\msun. The relation between the derived stellar mass by PARSEC and $M_{K_s}$ is displayed in Extended Data Fig.2.
 
The uncertainty of individual stellar mass was estimated via a Monte Carlo technique. Specifically, for each star, we calculated the stellar mass 1,000 times. Each time we drew a set of random values of $M_{K_s}$, $r_{\rm rest}$, and ${\rm T_{\rm eff}}$ from Gaussian distributions, with their measured values and uncertainties as to the mean and standard deviation, respectively. After the 1,000 time run, we obtain a probability density distribution (PDF) of the stellar mass. The resulting median value and standard deviation of each PDF were adopted as the derived stellar mass and the associated uncertainty, respectively. The typical random error (without systematic error) of the stellar mass is $\sim0.01$\,\msun. We note that the systematic error may increase the estimates by 0.01 \msun\, due to the stellar model.

We further compared the derived stellar mass with those determined by the empirical mass-luminosity relation (MLR)\cite{2019ApJ...871...63M}, which was benchmarked using orbital parameters of 62 binary stars within a stellar mass range of 0.075--0.70\,\msun. The mean value of the difference is only 0.01\,\msun\ with a 0.02\,\msun\ dispersion, which will not alter our results dramatically. 
%meaning that our stellar mass estimates are 
%which confirms that the stellar mass estimates are of both high accuracy and precision. 

In this work, we adopted a single power-law function to describe the IMF in the mass range of $0.3$--$0.7$\,\msun. This is because both canonical IMFs (Kroupa and Chabrier) may not be optimal for this mass range. The Kroupa IMF\cite{kroupa2001} is a segmented power-law function, with two broken points at 0.5 \msun\, and 1 \msun\, motivated by observation data. Our mass range goes across the discontinuity at 0.5 \msun, which is not driven by physics\cite{1979ApJS...41..513M}. The Chabrier IMF adopts a log-normal function\cite{chabrier2003} with a characteristic mass ($m_{\rm c}$) range of 0.15--0.25 \msun\cite{chabrier2003}. To avoid the degeneracy between $m_c$ and $\sigma$ of the log-normal function, a single power-law is more favoured when the focused mass range is larger than the characteristic mass\cite{2017MNRAS.468..319E}. 
%Our results (see Fig~2 a) also show that this stellar mass range of IMF is adequately described using a single power-law formula.
 
\subsection*{Stellar density profiles}
We selected the M-dwarf stars in a distance range of 100--300 pc and an $M_{K_S}$ range of 4.5--7.0 mag to avoid the Malmquist bias during star counting in the flux-limited sample. After the selection, the number of stars is uniformly distributed in the parameter space of distance, [M/H], and \msun\, (see Fig~1). The final selected sample contains 93,159 M-dwarf stars. 
  
The selection bias, induced by the targeting strategy design, the observation conditions, data process of the LAMOST survey, affects the number density profile measurements. We chose a Bayesian statistical method\cite{liu2017}, which was designed for LAMOST to derive the stellar density of each star for selection effect correction. 

Given a plate (a 5-degree-diameter sky area that LAMOST covers in an observation) with the central position in Galactic coordinates, ($l$, $b$), the targets that LAMOST successfully observed with a reasonable signal-to-noise ratio can be considered as a selection from a complete photometric catalogue (we use 2MASS catalogue in this work) in the same sky area. We note that the 2MASS catalogue has a completeness of 99\% when $K_s < 14.3$ mag\cite{2006AJ....131.1163S}, and 100\% of the considered sample are brighter than 14.3 mag. These targets are distributed in a colour-magnitude diagram. The selection function $S(mag,c|l,b)$ can be defined by the ratio of the number of targets to that of the photometrically completed samples at each point of magnitude and colour, ($mag$, $c$). The stellar density of the spectroscopic samples at a distance $D$ in the plate can be obtained by 
\begin{equation}
    \nu_{sp}(D|mag, c, l, b)=\nu_{ph}(D|mag, c, l,b)S(mag,c,l,b),
\end{equation}
where $\nu_{sp}$ and $\nu_{ph}$ are the stellar density of the spectroscopically observed stars, which can be obtained directly from observation, and that of the photometrically completed stars, which should be the true density to be determined, respectively. By integrating over colours and magnitudes, we have
\begin{equation}
    \nu_{ph}(D|l,b)=\iint\nu_{sp}(D|mag, c, l, b)S^{-1}(mag,c,l,b) \mathrm{d}mag \mathrm{d}c.
\end{equation}
$\nu_{sp}$ can be derived in terms of kernel density estimation by considering the probability distribution of the distances of stars observed in the plate of interest, that is
\begin{equation}\label{eq:nusp}
    \nu_{sp}(D|l,b)=\frac{1}{\Omega D^2}\sum_{i}{p_i(D)},
\end{equation}
where $p_i(D)$ represents for the probability distribution of distance of the $i$th star in the plate, which has a solid angle $\Omega$. We adopt the posterior distribution of distance\cite{2018AJ....156...58B} (via 15 and 85 percentiles) to construct p(D) of individual stars by combining two half-probability distributions. The left part of p(D) is a left half-Gaussian distribution with 15 percentile of distance as the sigma, while the right part of p(D) is the right half-Gaussian distribution with 85 percentile of distance as the sigma. Therefore, p(D) can be skewed since, in most cases, 15 and 85 percentiles are not symmetric. Ideally, $\nu_{sp}(D|l,b)$ is a continuous function of $D$. However, to avoid large uncertainty at positions with no star sample, we practically select discrete values of $\nu_{sp}(D_i|l,b)$, where $D_i$ is the most likely distance of the $i$th star, to involve in the calculation of stellar density profiles, as seen in Eq~(\ref{eq:expdisc}).

% In principle, Eq~(\ref{eq:nusp}) leads to an over-smoothing of the density, since it is a Gaussian convolution of the error-smoothed distribution. However, 
The selection correction method has been validated using simulated data, and it is found that it can well reproduce the logarithmic stellar density at the location of an individual star with uncertainty around 0.5\cite{liu2017}. With multiple stars in the same line-of-sight and multiple plates overlapped in the same line-of-sight, the averaged stellar density is usually very robust and only leads to at most 13\% uncertainty in the scale height estimates for a vertical exponential disc\cite{liu2017}. Besides, this method has been proved to work pretty well in several published studies on the structures of the Milky Way in the last five years.
%First, we assume that the 2MASS catalogue is a complete sample for stars with $K_s < 14.3$ mag. Second, in the colour-magnitude plane, we retrieve a selection function at each line of sight by modelling a number distribution based on 2MASS photometry\cite{liu2017}. Last, we obtain the intrinsic stellar density profile by multiplying this selection function with the stellar density profile observed by LAMOST. 

 \subsection*{Hierarchical Bayesian model}

We employed a hierarchical Bayesian model to derive IMF parameters by taking the stellar number density profile of the Galactic disc into account. We modeled the vertical stellar density with an exponential profile of the thin disc\cite{2008ApJ...673..864J}, which dominates the stellar distribution within 300 pc in the Solar neighbourhood. We neglect the stellar components of the thick disc and the halo. The derived profile model is shown as the following equation:

\begin{equation}\label{eq:expdisc}
    \nu_i(z_i \mid m_i) = \nu_0(z_i \mid m_i) \cdot exp(-\frac{\mid z_i \mid}{h_0}),
\end{equation}
where $\nu_i(z_i \mid m_i)$ is the spectroscopical number density ($\nu_{sp}$) of the $i^{\rm th}$ star with a stellar mass of $m_i$, $z_i$ is the height above/beneath the mid-plane in the cylindrical coordinate system of the Galactic disc ( 0 pc $<z_i<$ 300 pc), $h_0$ is the scale height, and $\nu_0(z_i \mid m_i)$ is the number density at $z_i = 0$. Here we assume that the IMF of a given metallicity does not vary with $z_i$ and $h_0$ with stellar mass in the studied volume.

% \sout{To optimise the completeness, we use M-dwarf stars in a mass range of $0.4$--$0.6$ $M_{\odot}$ to derive the thin-disc prior. The resulted Gaussian profile is centred at 512 pc with a standard deviation of 25 pc ($p(h_0) \propto \mathcal{N}(512, 25)$).}

%derived from M-dwarf stars with
%$0.4-0.6$\msun.

We then describe the IMF using a single power-law function with a power exponent \al\ and a constant $C$, 

\begin{equation}\label{eq:imf}
    \begin{aligned}
        \frac{\mathrm{d} n}{\mathrm{d} m} = C \cdot m^{-\alpha}.
    \end{aligned}
\end{equation}

The joint posterior distribution of the IMF parameters ($\alpha$ and $C$) and
the scale height $h_0$ can be written as: 

\begin{equation}\label{eq:post}
    p(\alpha, C, h_0 \mid {m_i, \nu_i, z_i}) \propto p(\alpha, C) \cdot p(h_0) \cdot \mathcal{L}(\{\nu_i\} \mid \{m_i, z_i\}, \alpha, h_0, C).
\end{equation}

We set flat prior with uniform distributions from 0 to 4, 3 to 7, and 100 to 1000, respectively, for \al\, $\log C$ and $h_0$. 
% \sout{For $h_0$ we adopt the Gaussian prior derived from Eq \ref{eq:expdisc}}. 
Then, we integrated the exponential density profile over $z$. The total derived number of stars in the range of stellar mass from $m$ to $m+\mathrm{d} m$ can be written as

\begin{equation}\label{eq:number}
    \begin{aligned}
    \frac{\mathrm{d} n}{\mathrm{d} m}  = \int_0^{\infty} \nu_0(m) \cdot exp(-\lvert {z} \rvert/h_0) \mathrm{d} z = \nu_0(m) h_0.
    \end{aligned}
\end{equation}

Combining Eqs~(\ref{eq:imf}) with~(\ref{eq:number}), we get 

\begin{equation}\label{eq:nu0}
    \begin{aligned}
    \nu_0(m)  = \frac{C \cdot m^{-\alpha}}{h_0},
    \end{aligned}
\end{equation}

We then brought Eq~(\ref{eq:nu0}) back to Eq~(\ref{eq:expdisc}) and obtained
the number density of each star:

\begin{equation}\label{eq:lnnu}
    \nu_{\mathrm{model},i} = \frac{C \cdot m_i^{-\alpha} \cdot exp(-\lvert {z_i} \rvert/h_0)}{h_0}.
\end{equation}

Last, for $N$ observations of the spectroscopic stellar number density ($\nu_i$) with associated stellar mass ($m_i$), height ($z_i$), and uncertainty of number density ($\sigma_{\nu_i}$), the likelihood in Eq~(\ref{eq:post}) is defined as

\begin{equation}\label{eq:likelihood}
\begin{gathered}
\mathcal{L}(\{\nu_i\} \mid \{m_i, z_i\}, \alpha, h_0, C)=\prod_{1}^{N} \exp\left( -\frac{(\nu_{\mathrm{model}}(z_i \mid m_i, \alpha, C, h_0) - \nu_{ph}(D \mid z_i))^2}{2 \sigma_{\nu_i}^2}\right). 
\end{gathered}
\end{equation}

We separate M-dwarf stars into metallicity bins of $-1.2$, $-0.7$, $-0.5$, $-0.3$, $-0.2$, $-0.1$, $+0.0$, $+0.2$ and $+0.5$ dex, with bin widths of 0.5, 0.2, 0.2, 0.1, 0.1, 0.1 and 0.3 dex, respectively. The non-uniformity of the bin width is to ensure that at least 500 stars are enclosed in each bin. The minimum bin width is close to the typical uncertainty of [M/H], 0.12 dex, which would not significantly smear out the results of the neighbouring bins.

To obtain the posteriors of the IMF parameters, we adopted a Markov Chain Monte Carlo (MCMC) sampling using the {\tt\string PyMC3} software package\cite{salvatier2016probabilistic}. After marginalising $\log C$, the median value and 15\% (85\%) of the posterior distribution of \al\ were adopted as the IMF slope and its corresponding uncertainty. 

To verify whether the IMF depends on the exponential disc model, we further test the results by direct counting. This time we do not consider the disc distribution and only select a narrow scale-height range of $100<\lvert {z} \rvert<250$ pc, where the observations were less affected by selection effects. The likelihood of each star can be expressed with the spectroscopic number density $\nu_i$ at its location: 

\begin{equation}\label{eq:likelihood_simple}
\begin{gathered}
\mathcal{L_2}(\{\nu_i\} \mid \{m_i\}, \alpha, C)=\prod_{1}^{N} \exp -(\frac{(C \cdot {m_i}^{-\alpha} - \nu(D \mid z_i))^2}{2 \sigma_{\nu_i}^2}). 
\end{gathered}
\end{equation}

Although the direct counting method has a slightly larger uncertainty, it is consistent with the results from the Bayesian model with an exponential Galactic disc (see Extended Data Fig. 3a). The IMF slopes obtained from the two methods differ by a maximum of 0.09, less than the average 1-$\sigma$ uncertainty ($\sim 0.14$), indicating the robustness of our results.

\subsection*{Validation with mock data}
%The performance and the effect of the sample size for the hierarchical Bayesian model given a group of stars. We performed a series of tests with mock data to validate our method further.

We generate 3,563,892 mock stars with \emph{Galaxia}\cite{2011ApJ...730....3S}, which adopts $1.6$ as the IMF power-law slope. Stellar mass, magnitudes, distances, effective temperature, ages and metallicities are created automatically. Binary stars are not considered here. All mock stars have stellar mass $<1$\,\msun, and they are distributed in a 20-square degree area toward the northern Galactic pole.

We selected stars inside a 2-kpc radius, with an absolute magnitude range of 3 mag $< M_{K_s} < $ 14 mag. This simulated data does not have selection bias, so we do not need to correct their number density. We estimate the stellar density in a moving cube of $50 \times 50 \times 50 ~ pc^3$ for each star. It is worth to note here that the mock sample is not selected using the same magnitude and radial limit as the observation data because then we will get less data than observations. The main purpose of the test is to validate the Bayesian method. And for a complete sample, different choices will not make much difference. 

We randomly drew subsets of 500, 1,000, and 10,000 mock stars from the sample and derived the number density profile for each subset. Then we applied the same hierarchical Bayesian model to each subset of stars and obtained the posterior distribution of \al. We find that as long as a subset contains more than 1,000 mock stars, the accuracy of \al\ is $\lesssim$0.1 dex.

To further verify whether our method can correctly evaluate $\alpha$ in mass-dependent selection effects, we selected 32 sky fields within an area of 25 square degrees from the \textit{Galaxia}'s mock data. We simulated a selection effect mimicking the real spectroscopic survey, similar to the literature\cite{liu2017}. The detailed selection function is split into two parts at $K_S = 13$ mag: for $K_S < 13$ mag, arbitrary stars are selected, and for $K_S > 13$ mag, a flat selection function is applied so that the mock stars are uniformly distributed along $K_S$. The adopted selection function emulates the strategy where LAMOST targeting is biased towards brighter instead of faint objects. The mass distribution below 0.5 $M_{\odot}$ is flattened in the selected samples than in the complete samples. The power-law slope falls from the complete sample (with ground true $\alpha = 1.6$) to $\sim 0.8 \pm 0.3$ after applying the selection function. 

Given a hypothetical distance error, we estimate the density $\nu_{sp}$ of each star using the same method. To mimic real observations, we add Gaussian-distributed random noises to the inverse of the true distance (parallax) given a parallax over error.
We used the same Bayesian model presented in this work to obtain the posterior distributions of power-law slope $\alpha$. 
We made the following two tests: First, we added different relative errors of parallax to the mock data and found our method can give correct $\alpha$ and $h_z$ estimates within $1\sigma$ uncertainty in the sample that have been subjected to selection effects. When the sample data size was about 10,000, the artificial parallax error had little effect on the results, and the dominant errors were Poisson noise. 
Second, we tested the effect of Poisson noise on the results for different sample sizes. We see that even with a sample size of 1000, we still have correct $\alpha$ with larger random error, while the estimate of $log_{10} ~ h_z$ is off the ground truth by about $2\sigma$.

Fortunately, 5 of 8 metallicity group (median [M/H]$=-0.4$, $-0.25$, $-0.15$, $-0.05$ and $+0.05$) in our sample has stars more than 10,000. The sample sizes of [M/H]$=-0.8$ and $-0.6$ are larger than 5,000. The derived $\alpha$ and $h_z$ should be reliable based on the above tests. The group of [M/H]$=+0.25$ has stars of 1,444. In such case, $\alpha$ is still credible with a largely biased $h_z$ estimates according to the mock data tests.

\subsection*{Validation with other mass-luminosity relations}
The mass-luminosity relation (MLR) plays a crucial role in determining the stellar mass. We have additionally used both Dartmouth stellar model\cite{2008ApJS..178...89D} and empirical MLR from literature\cite{2019ApJ...871...63M} separately as external validations to the results derived by the PARSEC model. We use the same data-driven algorithm to derive the stellar mass of each star. Then we adopt the same Bayesian model to derive the posterior distribution of IMF slopes. 

As displayed in Extended Data Fig.3(b), although the results given by the Dartmouth stellar model and empirical MLR have slight systematic differences from the PARSEC model, they do not significantly change the results given by PARSEC. We can still see almost the same $\alpha$ variation with the metallicity.

\subsection*{Vertical actions and stellar ages} 

We took \textit{galpy}\cite{2015ApJS..216...29B} to measure the 3-D location of each star in heliocentric Cartesian coordinates with velocity components ($V_R, V_T, V_Z$) in the Galactocentric cylindrical coordinate. The 3-D velocities of each star were calculated using line-of-sight velocity from LAMOST, parallax from {\it Gaia}, and proper motion from {\it Gaia}. The position of the Sun was adopted as (X, Y, Z) = (8178, 0, 27) pc\cite{2019A&A...625L..10G}. We also adopted the velocity of the local standard of rest (LSR) with respect to the Galactic center as $V_{\rm LSR}$ = 220 ${\rm km \cdot s^{-1}}$\cite{2012ApJ...759..131B} and the Solar motion with respect to the LSR as ($U_{\odot}$, $V_{\odot}$, $W_{\odot}$) = (11.1, 12.24, 7.25) ${\rm km \cdot s^{-1}}$\cite{2010MNRAS.403.1829S}.  

The orbital actions, including the vertical action \jz, were measured in the angle-action coordinate \cite{2008gady.book.....B, 2012MNRAS.426.1324B} with \textit{galpy}. The actions were estimated under the St\"{a}ckel approximation \cite{2012MNRAS.426.1324B} with the Milky Way potential {\tt\string MWPotential2014}\cite{2015ApJS..216...29B}. In a static or quasi-static axisymmetric gravitational potential, the amplitudes of a star's oscillation parallel and perpendicular to the Galactic plane can be described by the action $J_r$ and \jz\cite{2008gady.book.....B}, respectively. The third action $J_{\phi} \equiv L_z$ \cite{2016MNRAS.457.2107S} stands for the angle momentum with respect to the $z$ axis ($L_z$). \jz\ is defined as

 \begin{equation}
     J_z = \frac{1}{2 \pi} \oint {\rm d}z v_z,
 \end{equation}

where $z$ and $v_z$ are the instantaneous vertical position and vertical velocity of a star, respectively.

In general, \jz\ is correlated with stellar age\cite{2019ApJ...878...21T} since it continuously increases by scattering with giant molecular clouds\cite{1990MNRAS.245..305J}. Although the \jz--age relation is theoretically explicit, the observed \jz--age relation has a large dispersion, making it impossible to determine the exact age for an individual star\cite{2019ApJ...878...21T}. Nevertheless, a large \jz\ means that the star has experienced more scattering and thus is more likely to be old. On the other hand, a small \jz\ does not necessarily mean that the star is young since a star may not be effectively scattered over a long time. Therefore, we can separate the older population by selecting stars with large \jz\ but not selecting young populations simply from \jz. 

We benchmark the old population using red giant stars\cite{2018MNRAS.475.3633W} whose age is calibrated by asteroseismology, with an uncertainty of $\sim24$\%. We selected dynamically-hot stars with \jz$>20$\,\kpckms, mostly older than 4 Gyr (peaking at $\sim$8.5 Gyr), as the probe of old populations. We also selected the dynamically-cold stars with \jz$<10$\,\kpckms\ for reference. Extended Data Fig~4 (a) and (b) show that the dynamically-hot red giant stars are averagely old, while the dynamically-cold stars broadly spread in all ranges of age. We apply the same \jz\ criteria obtained from red giants to low-mass main-sequence stars because the \jz--age relation would not significantly change between them. 

We find that the ratio between dynamically-hot and -cold stars decreases against metallicity (see Extended Data Fig.4c). An exception, however, is the metallicity bin of [M/H]$=+0.25$, which contains more dynamically-hot stars than at $-0.3<$[M/H]$<0.1$. This bin contains more old stars than its neighbours, consistent with previous findings that they were migrated from the inner disc\cite{2015MNRAS.447.3526K}.

\subsection*{The impact of binarity in stellar mass estimation}

The influence of binary stars is two-fold: a) the unidentified binaries may be overestimated in luminosities and consequently masses, and b) the two stars in a binary system should be all counted in IMF. Here we discuss the first issue and leave the second in the next section.

The stellar mass of a primary star of a binary system might be overestimated due to the increased magnitude contributed by the secondary star. The shift of the absolute magnitude of an unresolved binary star from the single sequence in Hertzsprung-Russel diagram reflects the mass ratio ($q = m_2/m_1$, where $m_1$ and $m_2$ are the masses of the primary and secondary stars, respectively) of the binary system. 

When $q>0.7$, the contribution to the luminosity from the secondary star is $>$ 0.1 mag \cite{liu2019}, higher than the typical photometric uncertainty of $\sim0.07$ mag for $M_G$ (absolute magnitude in Gaia $G$-band).

Therefore, the combined luminosity for such binaries cannot be neglected compared to single stars. In this scenario, we assume that 25\% of M-dwarfs are in binary systems\cite{2004ASPC..318..166D} and the secondary-to-primary mass ratio is larger than 0.7 for 30\% binary stars. These would significantly affect the observed luminosity by $>$ 0.1 mag. Therefore, the absolute magnitudes of $\sim 7.5\%$ stars in our sample are overestimated for more than 0.1 mag, which may lead to an excessive estimation of stellar mass.

We again adopt mock stars generated by {\it Galaxia} to measure \al\ shifted by the overestimated mass from unidentified binary stars. The hidden secondary stars were randomly simulated for given binary fractions and mass-ratio distributions (in a power-law form with a power exponent, $\Gamma$). The $M_{K_S}$ of the mock stars were calculated using the mass-luminosity relation calibrated by dynamical stellar mass\cite{2019ApJ...871...63M}. Moreover, the $M_{K_S}$ of the primary stars were reacquired by eliminating the flux contribution of the secondary stars. Then we recalculate the stellar mass of the primary stars following the same method as the LAMOST M-dwarf stars.

We simulated the mock stars with binary fractions from 15\% to 40\%. We found that the effect of the overestimated stellar mass is limited for the derivation of \al. The change in \al\ by binary stars is less than $\sim$ 0.1.

\subsection*{Binarity correction in IMF}

To accurately obtain the IMF, we correct both the binary companions in counting stars and the overestimated stellar masses of the primary stars.
% \sout{We adopted the binary fraction, $f_b$, in a range of 0.0, 0.25, 0.75 and, 1.0, and the mass ratio distribution of binary stars, $\Gamma$, in a range of 0.0, 0.5, 1.5, and 5.0, respectively. For each combination of $f_b$ and $\Gamma$, we randomly draw subsets of 500, 1,000, 10,000, and 100,000 mock stars. We then measure \al\ with the Bayesian model for each mock dataset and derive the deviation of \al, $\Delta \alpha$, from the ground truth.}
We adopted the binary fraction, $f_b$, in a range from 0 to 1. We randomly draw subsets of 1,000, 10,000 and 100,000 mock stars. In each simulation, a group of stars are generated following a given mass function ($\alpha$=1.6, the value used for \textit{Galaxia}) from 0.1 to 3.0 $M_{\odot}$. Then binary stars are selected from the whole sample by randomly paring. The binary fraction ($f_b$) is defined as
\begin{equation}
    f_b = \frac{N_{\rm bin}}{N_{\rm bin} + N_{\rm sig}},
\end{equation}
 where $N_{\rm bin}$ and $N_{\rm sig}$ are the number of binary systems and single-star, respectively. In each binary system, the secondary star is regarded as unresolved. All the unresolved secondary stars are then excluded, and leave the remaining as the single and primary stars in the group. We take the MLR\cite{2019ApJ...871...63M} to derive $M_{K,2}$ of secondary stars, and measure the combined $M_{K,{\rm tot}}$ of each binary system. Then we obtain stellar mass ($m_1'$) from $M_{K,{\rm tot}}$ using MLR\cite{2019ApJ...871...63M}. In this way, the primary mass of each binary system is slightly overestimated caused by the additional luminosity of the companion star. We then use the Bayesian model to measure the remaining stars' power-law slope $\alpha$. 

We generated binary stars randomly ten times in each dataset and obtained the mean $\Delta \alpha$ and its standard deviation. As shown in Extended Data Fig.5, $\Delta \alpha$ increases with binary fraction. $\Delta \alpha$ is about 0.7 when the binary fraction reaches 60\%, the same level as previous classical studies. When the binary fraction is 100\%, $\Delta \alpha$ is $\sim$ 1.0. 
% \sout{Even if all stars are in binary systems, the derivation of \al\ from the ground truth is less than 0.25$\pm0.1$ dex.} 
Given 30\% M-dwarfs are in binary systems, which is not far away from many observed values\cite{liu2019,2019ApJ...875...61M}, $\Delta \alpha$ is $\sim$ 0.3.
% \sout{, which would not significantly change our main results. }

%Previous binary studies\cite{2019ApJ...875...61M,liu2019} show binary fraction %and mass-ratio distribution at various metallicities and masses in the Solar %neighbourhood.%weighted by numbers in each bin of primary stellar mass ). 

We adopted the binary fractions at different [M/H] measured from stars in the Solar neighbourhood (Liu 2019, hereafter L19\cite{liu2019}; Moe et al. 2019 hereafter M19\cite{2019ApJ...875...61M}). Then, we interpolate the binary fraction to our [M/H] bins. The adopted values, the corresponding $f_b$ and corrected \al\ are listed in Extended Data Table 2. The binary corrected \al\ is drawn as a solid grey line and dotted line in Fig.2b.

%\lc{add a paragraph (mainly copy from reply to the referee) about that the extreme fbin-[M/H] slope also would not flatten the alpha-[M/H] relation?}

Now we consider the variation of the IMF of literature\cite{2020A&A...637A..68Y} (Y20) as a benchmark. Because the binary fraction and [M/H] are correlated, we used various $f_b$--[M/H] relations (${\rm d}f_b/{\rm d}{\rm [M/H]}$), in conjunction with the tests of $f_b$--$\Delta \alpha$ described above, to investigate whether the $f_b$--[M/H] gradient affects the $\alpha$--[M/H] relation. As shown in Extended Data Fig. 5, the slope of the anti-correlation between $f_b$ and metallicity is ${\rm d} f_b / {\rm d} {\rm [M/H]} \sim -0.12$ according to M19 and ${\rm d} f_b / {\rm d} {\rm [M/H]} \sim -0.20$ from the the group with high mass-ratio distribution binary stars of L19. Under the binary correction of these two studies, the variation of the slope of IMF is still significant, as displayed with blue and orange solid lines, respectively, in Extended Data Fig. 5. Even though we consider extremely high slopes of -0.5 and -1.0, which are significantly higher than all observational reports, the variation of $\alpha$ remains at around solar metallicity ($-0.2<$[M/H]$<-0.1$), although the IMF flattens in the regime of [M/H]$<-0.3$ (see Extended Data Fig. 5.). However, ${\rm d} f_b / {\rm d} {\rm [M/H]} = -1.00 $ is a unrealistic gradient, as if the binary fraction of [M/H]$=0$ is 0\%, the binary fraction of [M/H]$=-1$ is 100\%, which seems unreasonable. Therefore, the significant difference of $\alpha$ at different [M/H] should be a real signature unless the binary fraction of metal-poor stars is unrealistically high.

The dynamically-hot stars often have larger velocity dispersion, which may be contaminated by the additional orbital velocity in a close binary system. Therefore, we presume that the dynamically-hot stars may contain more binaries than the dynamically-cold stars. To quantify the binary fraction of the dynamically-hot stars, we compare the maximum radial velocity difference ($\Delta$RV$_{\rm max}$, the difference between the highest and the smallest radial velocities of a given star observed with multi-epoch observations.) distributions between the dynamically-hot and -cold stars. The peak of the $\Delta$RV$_{\rm max}$ distribution is dominated by the measurement errors of radial velocities, while the binaries mainly contribute to the distribution tail. Hence, $\Delta$RV$_{\rm max}$ between any two epochs for the same object can be used to detect binary stars\cite{2019ApJ...875...61M}. 

We select stars that have been observed for at least three epochs with LAMOST, which has a line-of-sight velocity uncertainty $\sim$ 11.5 ${\rm km \cdot s^{-1}}$\cite{2014AJ....147...33Y}. There are 2803 dynamically-hot and 108 dynamically-cold stars, respectively. Extended Data Fig.6 compares the $\Delta$RV$_{\rm max}$ distributions of the two groups. The stars with $\Delta$RV$_{\rm max}$ $>$ 30${\rm km \cdot s^{-1}}$ show noticeable the difference in distributions between the dynamically-hot and -cold groups. Dynamically-hot stars seem to contain more binaries. 
% \sout{We performed a Kolmogorov-Smirnov test obtained a p-value of 0.18, further illustrating that the difference between these two distributions is statistically significant. }

We assumed that stars with $\Delta {\rm RV}_{\rm max}>25{\rm km \cdot s^{-1}}$ (larger than 2-$\sigma$ of the typical uncertainty of radial velocity, which is 11.5 ${\rm km \cdot s^{-1}}$ of LAMOST M-dwarf stars\cite{2014AJ....147...33Y}) are primarily in binary systems. The fractions of these portions of stars are 0.15 and 0.1, respectively, for dynamically-hot and- cold samples. The binary fraction of the dynamically-hot group is a factor of $\sim$1.5 than that of the dynamically-cold group. Therefore, the binary fraction of the dynamically-hot population is $\sim45\%$, given that the average binary fraction of dynamically-cold stars is $30\%$ over the whole metallicity regime\cite{liu2019}. 

% \st{We note that the relatively high binary fraction of the dynamically-hot stars might be related to the findings that the distributed population of stars formed after the gas was expelled a higher proportion of binary stars than the stars remaining in the cluster.}%\cite{2001MNRAS.321..699K}.

\section*{Data availability}
The raw dataset that supports the findings of this study is publicly available at the National Astronomical Data Center (\url{https://nadc.china-vo.org/article/20200722160959?id=101069}). The data generated and/or analysed during the study are available at the National Astronomical Data Center \url{(http://paperdata.china-vo.org/jordan/jdli22_imf.csv)}.

\section*{Code availability}
The code used to determine the stellar mass of M-dwarf stars and model fitting is publicly available on Github at \url{https://github.com/jiadonglee/MDwarfMachine.}

\clearpage

%=====================End of main text===================================================
%========================================================================================
% For data citations of datasets uploaded to, e.g. \emph{figshare}, please use the \verb|howpublished| option in the bib entry to specify the platform and the link, as in the \verb|Hao:gidmaps:2014| example in the sample bibliography file.

\section*{Acknowledgements}
We thank the anonymous referees for their very constructive and helpful comments. We appreciate Prof. Licai Deng and Prof. Richard de Grijs for their contributions in the very early stage of the project. We thank Prof. Jifeng Liu for the helpful discussions. This work is supported by the National Key R\&D Program of China No. 2019YFA0405500. C.L. thanks the National Natural Science Foundation of China (NSFC) for grant No. 11835057 and the science research grants from the China Manned Space Project with NO.CMS-CSST-2021-A08. Z.Y.Z and Z.Q.Y acknowledge the support of NSFC grants No. 12041305 and No. 12173016, and the Program for Innovative Talents, Entrepreneur in Jiangsu. H.T. acknowledges the support of NSFC grant No.12103062. X.F. acknowledges the support of China Postdoctoral Science Foundation No. 2020M670023, the NSFC grant No. 11973001 and No. 12090044, and the National Key R\&D Program of China No. 2019YFA0405504. J.L. acknowledges the science research grants from the China Manned Space Project with NO.CMS-CSST-2021-A10 and NO.CMS-CSST-2021-B05, and the NSFC grant No. 12090043 and No. 11873016. Z.Q.Y acknowledges support from NSFC grants No. 12203021, the Jiangsu Funding Program for Excellent Postdoctoral Talent under grant number 2022ZB54, the Fundamental Research Funds for the Central Universities under grant number 0201/14380049. 
Guoshoujing Telescope (the Large Sky Area Multi-Object Fiber Spectroscopic Telescope, LAMOST) is a National Major Scientific Project built by the Chinese Academy of Sciences. The National Development and Reform Commission has provided funding for the project. LAMOST is operated and managed by the National Astronomical Observatories, Chinese Academy of Sciences. This work has made use of data from the European Space Agency (ESA) mission Gaia (https://www.cosmos.esa.int/gaia), processed by the Gaia Data Processing and Analysis Consortium (DPAC; https://www.cosmos.esa.int/web/gaia/dpac/consortium). Funding for the DPAC has been provided by national institutions, in particular the institutions participating in the Gaia Multilateral Agreement.
This work also benefited from the International Space Science Institute (ISSI/ISSI-BJ) in Bern and Beijing, thanks to the funding of the team “Chemical abundances in the ISM: the litmus test of stellar IMF variations in galaxies across cosmic time” (Principal Investigator D.R. and Z-Y.Z.) 

\section*{Author contributions statement}
% Must include all authors, identified by initials, for example:
% A.A. conceived the experiment(s),  A.A. and B.A. conducted the experiment(s), C.A. and D.A. analysed the results. All authors reviewed the manuscript.
Jiadong L. contributed most of the modelling and calculations and wrote the initial manuscript. C.L. provided the ideas to initialise the project, supervised Jiadong L. on the modelling, and revised the manuscript. Z.-Y.Z., X.F. and Z.-Q.Y. compared the results with other theoretical and observational work and helped write the manuscript. H.T. ran the calculations to derive vertical actions. J.L. helped with the discussion of the impact of the binary stars. All authors discussed and commented on the manuscript.

\section*{Author information}
Corresponding author: Chao Liu

\section*{Competing interest}
The authors declare no competing interests.

\section*{Additional information}
%To include, in this order: \textbf{Accession codes} (where applicable); \textbf{Competing interests} (mandatory statement). 
%The corresponding author is responsible for submitting a \href{http://www.nature.com/srep/policies/index.html#competing}{competing interests statement} on behalf of all authors of the paper. This statement must be included in the submitted article file.

\section*{Extended data}
\clearpage

\begin{figure*}
\centering
\includegraphics[width=0.9\linewidth]
{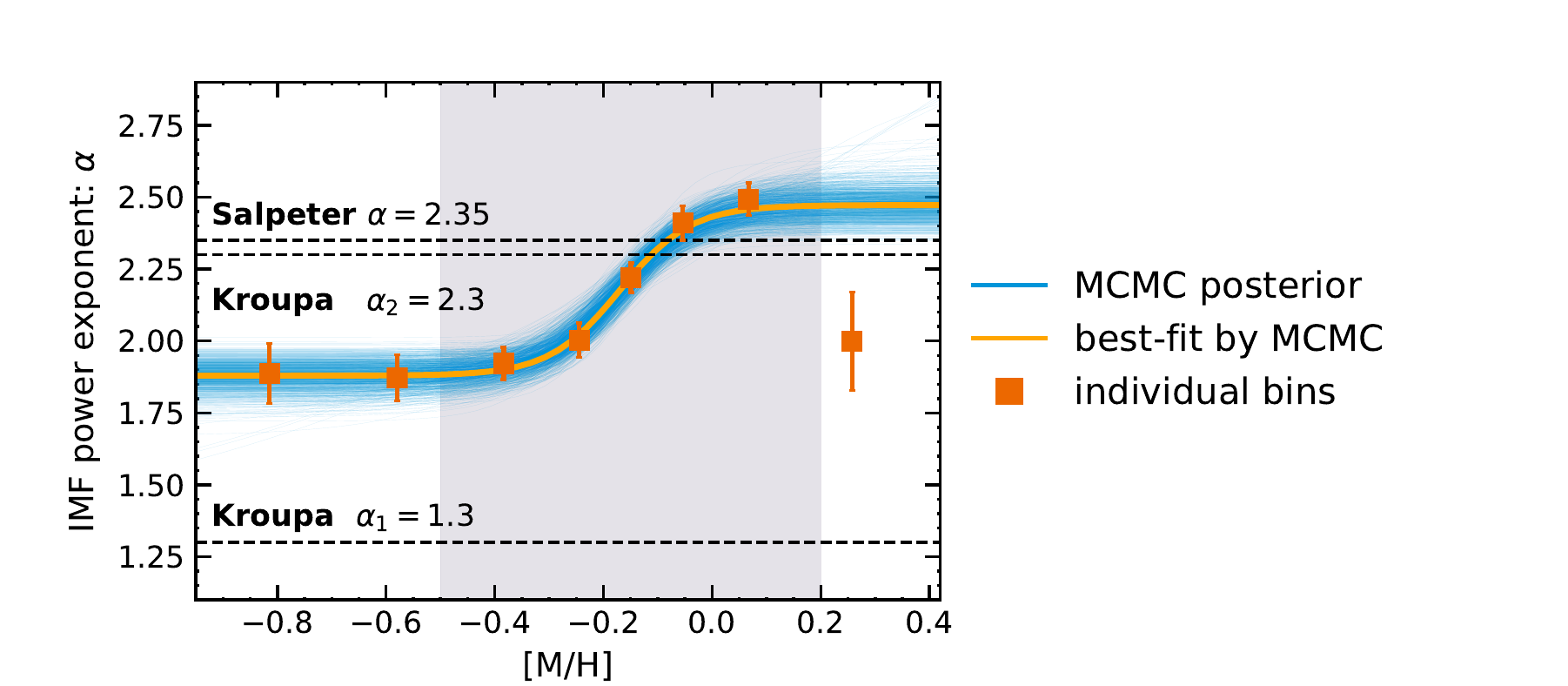}
\caption{\textbf{Extended Data Fig 1. The Sigmoid function fitting result of IMF power-law slope as a function of [M/H].} The yellow line is the best-fit result. The red squares are results in our study using to fit. The blue lines are 100 fitting results selected randomly from the MCMC chains. Error bars represent 1$\sigma$ uncertainty.}
\end{figure*}

\begin{figure*}
\centering
\includegraphics[width=0.7\linewidth]
{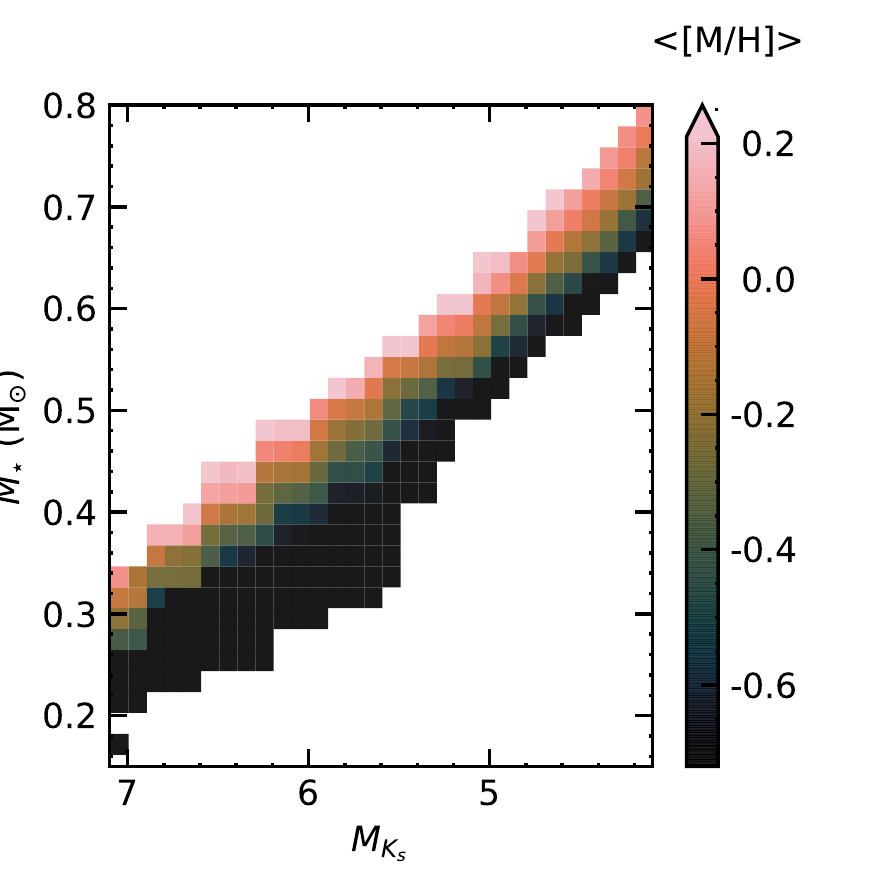}
\caption{\textbf{Extended Data Fig 2: The derived stellar mass by PARSEC model as a function of absolute magnitude of $K_S$-band ($M_{K_S}$).} The colours of each pixel represent the median [M/H] in each colour-magnitude bin.}
\end{figure*}

\begin{figure*}
\centering
\includegraphics[width=0.99\linewidth]
{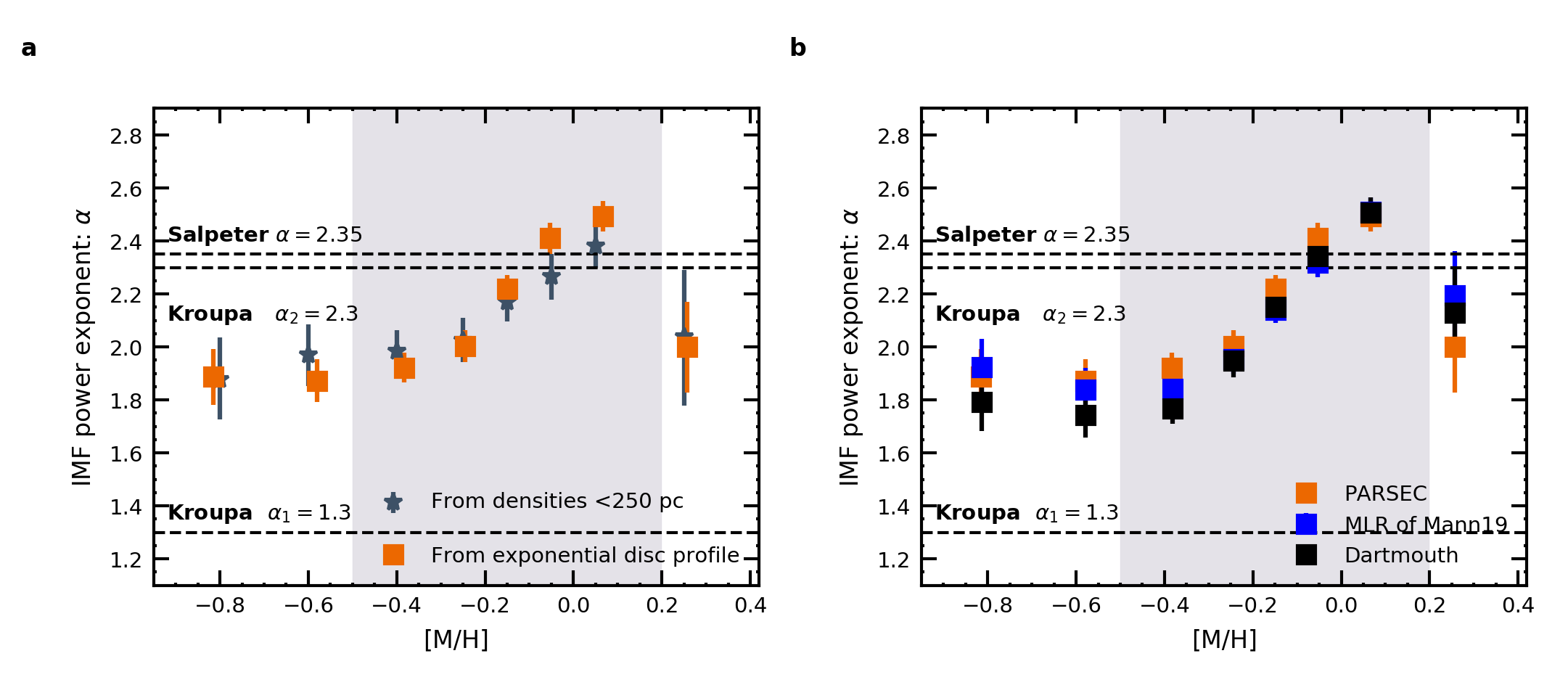}
\caption{\textbf{Extended Data Fig 3: The IMF power-law index as a function of [M/H] measured with different methods and different stellar models.} (a) The red filled rectangles are derived by all stars with the hierarchical Bayesian model considering the exponential Galactic disc profile. The filled stars denote the \al\,s derived directly from the observed densities of stars in the $\lvert{z}\rvert$ range from 100 to 250 pc. (b) The IMF power-law slope as a function of [M/H] based on various stellar models and empirical MLR, respectively. The red filled rectangles are derived by PARSEC, and the black rectangles are derived by Dartmouth\cite{2008ApJS..178...89D}. The blue rectangles denote $\alpha$s from the empirical mass-luminosity relation\cite{2019ApJ...871...63M}. Error bars represent 1$\sigma$ uncertainty.}
\end{figure*}

\begin{figure*}
\centering
\includegraphics[width=0.99\linewidth]
{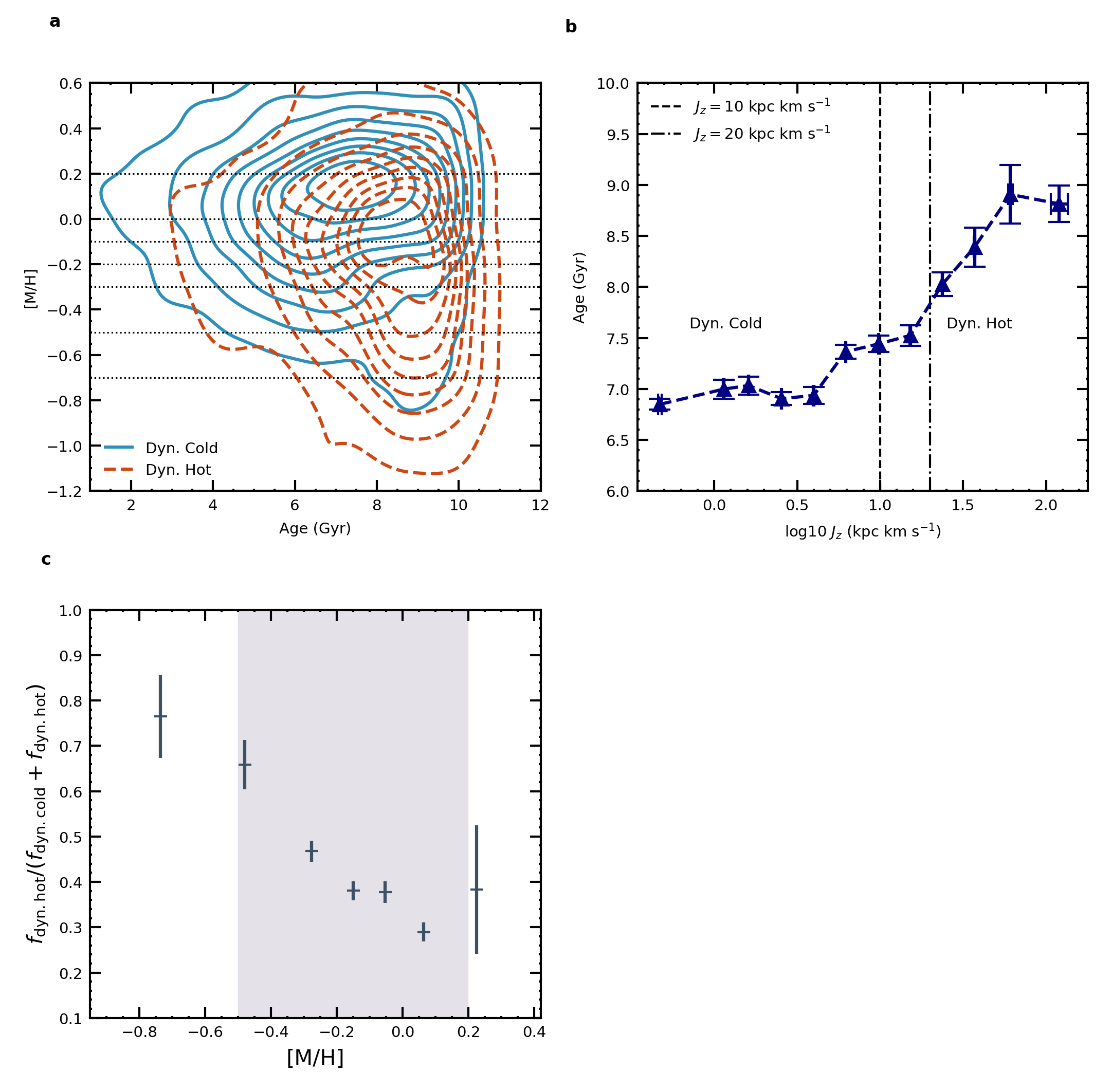}
\caption{\textbf{Extended Data Fig 4: Red giant branch stars\cite{2018MNRAS.475.3633W} reveals vertical actions \jz\ increase with stellar age.} (a) Age-metallicity distributions of red giant branch stars with in 500\,pc in the solar vicinity. The red and blue contours display the age-metallicity distribution of dynamically-hot (\jz$>20$\,\kpckms) and -cold (\jz$<10$\,\kpckms) stars, respectively. Their contour densities are smoothed by the kernel density estimation method. The dotted lines indicate the separation of [M/H] slices used in the M dwarf samples. (b) Stellar ages as a function of \jz\ in logarithmic form. The left-side of the vertical dashed line denotes the dynamically-cold stars, and the right-side of the vertical dash-dotted line represents the dynamically-hot stars. (c) The ratio of the normalised number of dynamically-hot stars to that of the dynamically-cold stars as a function of [M/H]. $f_{\rm dyn-hot}$ is the proportion of the number of dynamically-hot stars in each [M/H] bin to all dynamically-hot stars of M dwarf star sample. $f_{\rm dyn-cold}$ denotes the similar proportion, but for dynamically-cold stars. $f_{\rm dyn-hot} / (f_{\rm dyn-hot}+f_{\rm dyn-cold})$ indicate the normalised number ratio between dynamically-hot and the sum of dynamically-hot and -cold ratios. The vertical shaded region represents the metallicity range of $-0.5<$[M/H]$<0.2$, corresponding to the area of \al\ variation in Fig.2b and Fig.3. Error bars represent 1$\sigma$ uncertainty.}
\end{figure*}

\begin{figure*}
\centering
\includegraphics[width=0.99\linewidth]
{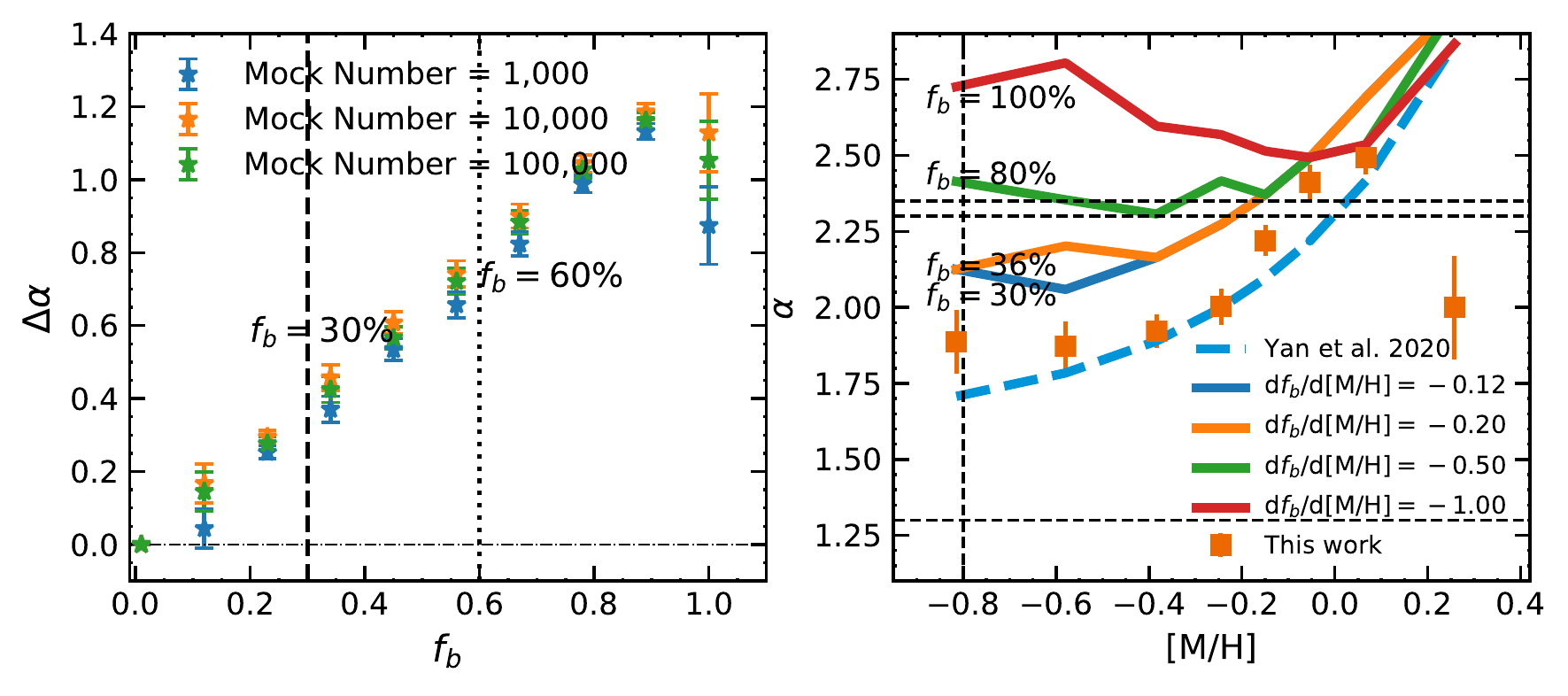}
\caption{\textbf{Extended Data Fig 5. The test results of the impact of the binary stars on the IMF}. (a) The results of simulations to verify the impact of the binary stars in the IMF. The results show the difference of the estimated $\alpha$ from the true values versus binary fraction with the numbers of mock stars equal to 1,000, 10,000 and 100,000. The vertical dotted line represents the binary fraction is 30\%, which is approximately the observed mean binary fraction for solar metallicity stars. The vertical dashed line denotes the binary fraction is 60\%. (b) The test results by setting different slopes of binary fraction as a function of [M/H]. The blue, orange, green, and red solid lines show the trend of $\alpha$ with [M/H] by adopting ${\rm d} f_b / {\rm [M/H]} \sim -0.12$, $-0.20$, $-0.50$, and $-1.00$, respectively. The blue dashed line indicates the IMF formula in Yan et al. (2020). The annotations on the right of the vertical line denote the binary fractions of [M/H]$=-0.8$. Error bars represent 1$\sigma$ uncertainty.}
\end{figure*}

\begin{figure*}
\centering
\includegraphics[width=0.7\linewidth]
{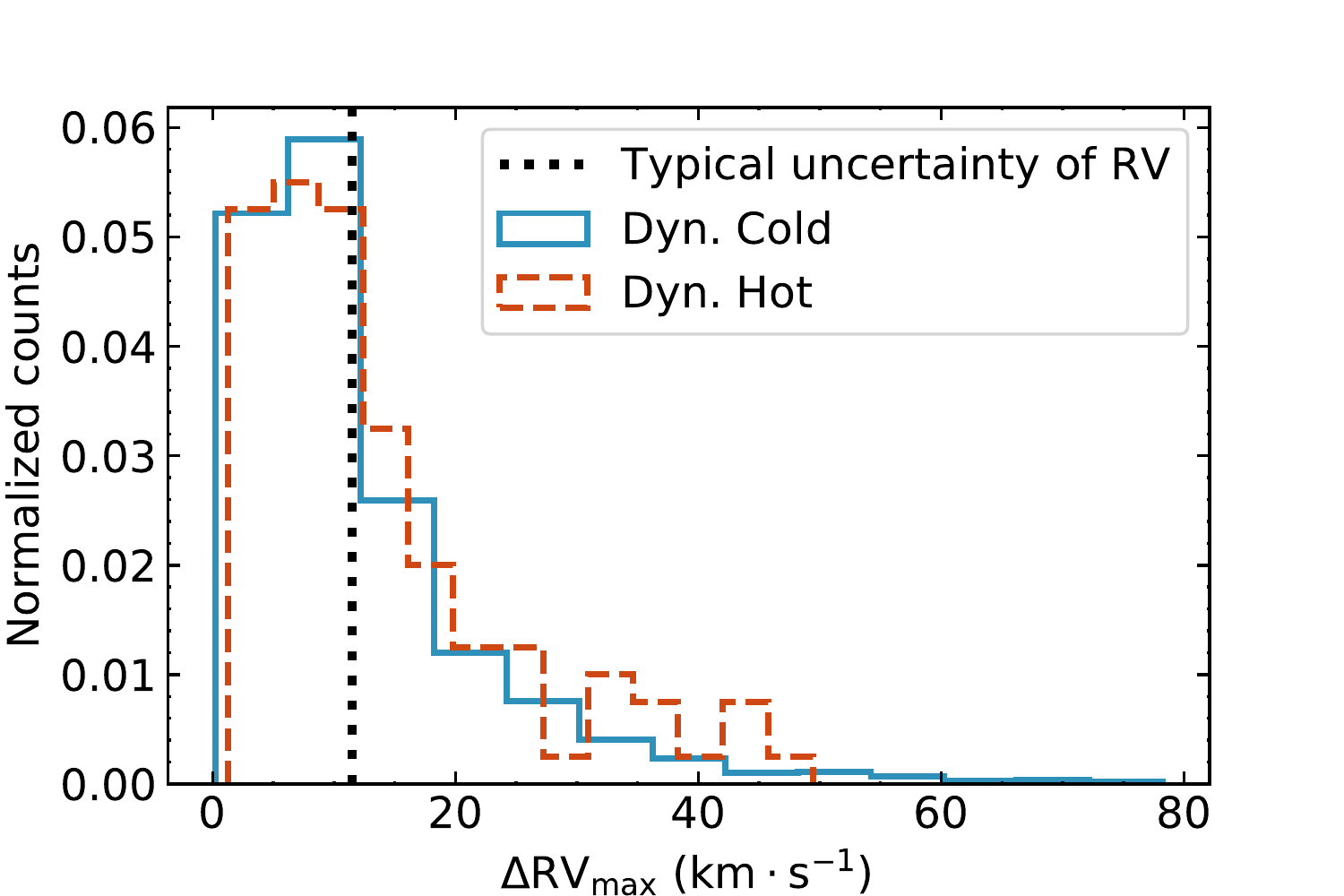}
\caption{\textbf{Extended Data Fig 6. The $\Delta$RV$_{\rm max}$ distribution of the dynamically-hot M dwarf stars (red solid-line), and dynamically-cold stars (blue dashed-line).} The vertical dotted line denotes the typical 1 $-\sigma$ uncertainty of radial velocity uncertainty of LAMOST M dwarf stars.}
\end{figure*}

% \textbf{Extended Data Table 1: The Sigmoid function fitting parameters of IMF power-law slope as a function and corresponding errors.}
% The Sigmoid function is $\alpha = \displaystyle{\frac{L}{1+e^{-k \cdot ({\rm [M/H]} - x_0)}} + \alpha_0}$.

\begin{table*}
\caption{\textbf{Extended Data Table 1:} The Sigmoid function ($\alpha = \displaystyle{\frac{L}{1+e^{-k \cdot ({\rm [M/H]} - x_0)}} + \alpha_0}$) fitting parameters of IMF power-law slope as a function and corresponding errors.}
\begin{tabular}{lll}\label{tab:mrich}
\\
\hline
\hline
Fitting parameters &  best-fitting value & uncertainty \\ \hline
L                  & 0.59  & 0.09     \\
$x_0$              & -0.17 & 0.03     \\
k                  & 15.43 & 3.36     \\
$\alpha_0$         & 1.88  & 0.05     \\ \hline
\end{tabular}
\end{table*}

% \textbf{Extended Data Table 2: Binary corrections to IMF slopes in different [M/H].} 
% The binary results are adopted from Liu 2019 (Liu19)\cite{liu2019} and Moe et al. 2019 (M19)\cite{2019ApJ...875...61M}, respectively.

\begin{table*}[]
\caption{\textbf{Extended Data Table 2:} Binary corrections to IMF slopes in different [M/H] using binary results from Liu 2019 (Liu19)\cite{liu2019} and Moe et al. 2019 (M19)\cite{2019ApJ...875...61M}, respectively.}
\begin{tabular}{l|c|c|c|c}
\hline
{[}M/H{]} & $f_b$ (Liu19) & $\Delta \alpha$ (Liu19)      & $f_b$ (Moe 19) & $\Delta \alpha$ (Moe 19)     \\ \hline
$-0.80$     & 0.30          & 0.42 $\pm$ 0.04              & 0.36           & 0.42 $\pm$ 0.04 \\ 
$-0.60$  & 0.30          & 0.42 $\pm$ 0.04              & 0.32           & 0.42 $\pm$ 0.04 \\ 
$-0.40$ & 0.30          & 0.42 $\pm$ 0.04              & 0.28           & 0.27 $\pm$ 0.03 \\ 
$-0.25$     & 0.33          & 0.42 $\pm$ 0.04              & 0.25           & 0.27 $\pm$ 0.03 \\ 
$-0.15$     & 0.33          & 0.42 $\pm$ 0.04 & 0.23           & 0.27 $\pm$ 0.03 \\ 
$-0.05$     & 0.29          & 0.27 $\pm$ 0.03 & 0.21           & 0.27 $\pm$ 0.03 \\ 
$+0.05$     & 0.23          & 0.27 $\pm$ 0.03 & 0.18           & 0.27 $\pm$ 0.03 \\ 
$+0.25$    & 0.24          & 0.27 $\pm$ 0.03 & 0.15           & 0.12 $\pm$ 0.03 \\ \hline
\end{tabular}
\end{table*}

\end{document}